\documentclass[twocolumn,prl,showpacs,superscriptaddress]{revtex4}
\usepackage{psfrag}
\usepackage{graphicx}
\usepackage{amsmath}
\usepackage{amssymb}

\begin{document}
\title{Renormalization of Optical Excitations in Molecules near a Metal Surface}
\author{J. M. Garcia-Lastra}
\affiliation{Center for Atomic-scale Materials Design,
  Department of Physics, Technical University of Denmark,DK - 2800
  Kgs. Lyngby, Denmark.}
\author{K. S. Thygesen}
\email[]{thygesen@fysik.dtu.dk}
\affiliation{Center for Atomic-scale Materials Design,
  Department of Physics, Technical University of Denmark,DK - 2800
  Kgs. Lyngby, Denmark.}

\date{\today}

\begin{abstract}
  The lowest electronic excitations of benzene and a set
  of donor-acceptor molecular complexes are calculated for the gas phase and on the
Al(111) surface using the many-body Bethe-Salpeter
  equation (BSE). The energy of the charge-transfer excitations obtained for the gas
phase complexes are found to be around 10\% lower than the experimental values.
When the molecules are placed outside the surface, the enhanced screening from the
metal reduces the exciton binding energies by several eVs and the transition
energies by up to 1 eV depending on the size of the transition-generated dipole. 
As a striking
  consequence we find that close to the metal surface the optical gap of benzene can
exceed its
  quasiparticle gap. A classical
  image charge model for the screened Coulomb interaction
  can account for all these effects which, on the other hand, are
  completely missed by standard time-dependent density functional theory.
\end{abstract}

\maketitle

The performance of organic-based (opto-)electronics devices such as
organic and dye sensitized solar cells, organic
transistors and light emitting
diodes\cite{bredas,graetzel,hwang}, depends crucially
on the interface between the ``active'' organic region and the
metallic or semiconducting electrodes. In particular, the position of the molecular
energy levels relative to the metal Fermi level, and the size of the electron-hole
binding energy are of key importance for the charge transport across an
organic-metal interface and for the dissociation of photoexcited excitons at a
donor-acceptor interface. The purpose of this Letter is to illustrate how the
excitation spectrum
of a molecule is affected by the presence of a nearby metal surface, see Fig. 1.

Elementary excitations of many-electron systems consist two basic types: (i)
Quasiparticle (QP)
excitations which involve the addition or removal of an electron from
the system, and (ii) optical excitations which involve the promotion
of an electron from one electronic level to another. In the case of molecules on
metal surfaces the former type can be probed in
transport- or photoemission
experiments\cite{repp05,himpsel} while the latter can be probed by electron
energy loss- or surface-enhanced Raman
spectroscopy\cite{avouris,moskovits,nie}.  It has recently been demonstrated both
experimentally\cite{repp05,kahn} and on the basis of
many-body calculations\cite{neaton,thygesen_image,garcia}, that when a
molecule is adsorbed on a polarizable substrate its QP gap, i.e. the difference between
ionization potential ($I_p$) and electron affinity ($E_a$), is reduced as a
result of image charge interactions. The same mechanism was shown
earlier to lead to a narrowing of the band gap at semiconductor-metal
interfaces\cite{inkson}. In contrast, similar effects on
the optical excitations of adsorbed molecules have so far only been
discussed on the basis of phenomenological
models\cite{chance,maniv,holland}.

An efficient method for the calculation of optical excitations in molecules is
the time-dependent density functional theory
(TDDFT)\cite{marques}. The widely used adiabatic local
density approximation (ALDA) has been quite successfull for
intramolecular transitions whereas charge-transfer excitations are
significantly underestimated.\cite{head-gordon} Alternatively, optical
excitations can be obtained from the Bethe-Salpeter equation (BSE)
which is rooted in many-body perturbation theory\cite{rubio_RMP}. This
approach has been successfully used to describe the optical properties
of a broad variety of systems including organic and inorganic
semiconductors as well as gas-phase
molecules\cite{rubio_RMP,BSE_benzene}. However, applications of the BSE to
charge-transfer complexes or
metal-molecule interfaces have so far not been reported.

\begin{figure}[!b]
\begin{center}
\includegraphics[width=0.55\linewidth]{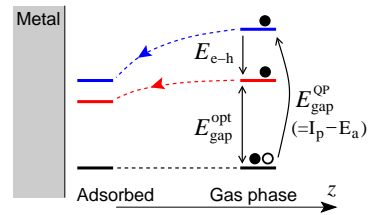}
\caption[system]{\label{fig1} (Color online) Schematic illustration of the narrowing
of the energy gaps of a molecule as it approaches a metal surface. The reduction of
the QP gap is due to the interaction of the added electron/hole with its image
charge in the metal. Similarly the optical gap is reduced due to the interaction of
the transition dipole with its induced image dipole. Finally, the exciton binding
energy ($E_{\text{e-h}}$) is also reduced because the electron-hole interaction is
screened by the metal.}
\end{center}
\end{figure}

In the first part of this Letter we assess the performance of the BSE
approach for describing charge-transfer excitations in a set of four
donor-TCNE (tetracyanoethylene) molecular complexes. In the second and
main part we investigate how the optical properties of a molecule
(here represented by benzene and a benzene-TCNE dimer) change when placed outside an
Al(111) surface.

\begin{widetext}
\begin{center}
\begin{table}[!t]
\caption{Experimental and calculated energy gaps of the four donor-TCNE complexes in
the gas phase. The experimental (calculated) QP gaps have been obtained as the
difference between the ionization potential of the free donor (the HOMO level) and
electron affinity of free TCNE (the LUMO level). The optical gap is the
corresponding $\pi$(donor) $\to$ $\pi^*$(TCNE) singlet transition of the complex.}
\begin{tabular}{l|ccc|cccc|ccc}

\hline \hline
          &  &  $E_{\text{gap}}^{\text{QP}}$ (eV)&   &  &
$E_{\text{gap}}^{\text{opt}}$ (eV) & & & &  $E_{\text{e-h}}$ (eV)  & \\

\hline
Donor         &\ G$_0$W$_0$   &  $\Delta$SCF     &$^a$Exp. & BSE & TDLDA &
$\Delta$SCF   &$^b$Exp.  & G$_0$W$_0$-BSE &  $\Delta$SCF   &$^{a,b}$Exp. \\

  \hline \hline
  Benzene        & 6.31 &    6.07  & 6.33 &  3.22 & 1.47 &  3.49   &                
  3.59  &                              -3.09 &   -2.58 & -2.74 \\
  Toluene        & 5.81 &     5.50  & 5.92 &  2.82 & 1.28 &  2.87   &               
   3.36  &                             -2.99 &   -2.63 & -2.56 \\
  O-xylene       & 5.59 &     5.21  & 5.65 &  2.73 & 1.42 &  2.57   &               
   3.15  &                             -2.87 &   -2.64 & -2.50 \\
  Naphthalene     & 5.04 &   4.68  & 5.23 &  2.38 & 1.93 &  2.12   &                
  2.60 &                              -2.66 &   -2.56 & -2.63 \\
  \hline
  MAE          & 0.11 &    0.42  & 0.0 & 0.39 & 1.65 &  0.42   &                  
0.0          &                               0.30 &  0.11 & 0.0 \\
  \hline \hline

\end{tabular}

\flushleft
$^{a}$Ionization potentials from Refs. \onlinecite{dewar,watanabe}. Electron
affinity of TCNE from Refs. \onlinecite{TCNE_lumo1,TCNE_lumo2} (we used the value
2.91 eV). \\
$^{b}$From Ref. \onlinecite{hanazaki}\\
\end{table}
\end{center}
\end{widetext}

The gas-phase calculations were performed in a 15 \AA~cubic supercell
with the donor-TCNE dimers fixed in the structures reported in Ref.
\onlinecite{stein_kronik}. The
donor-TCNE distance ($d$) lie in the range $3.57-3.91$ \AA~for the
four complexes. For the metal-molecule interfaces we used a supercell
containing three layers of Al(111) with $5\times 5$ atoms in each layer. The
molecules in their gas-phase structure were placed a distance $z$ above
the surface followed by 13 \AA~of vacuum. DFT-LDA calculations were performed with
the PWSCF
code\cite{pwscf} using a 40 Hartree
plane-wave cut-off and a $2\times 2\times 1$ $k$-point mesh (only the $\Gamma$-point
was used for the gas-phase calculations).  G$_0$W$_0$ and BSE calculations were
performed with the Yambo code\cite{yambo}
using the LDA wave functions and eigenvalues as input and a
plasmon pole model fitted to the dielectric matrix at imaginary frequencies 0 and 1
Hartree. For the dielectric matrix we used a 30 eV cut-off for the
sum over virtual states.  We have verified that increasing the number
of $k$-points in the surface plane to from $2\times 2$ to $4\times 4$
changes the G$_0$W$_0$ and BSE energies by less than 0.05 eV. The G$_0$W$_0$ and BSE
supercell calculations for the isolated molecules have been performed with a
truncated Coulomb interaction in order to avoid spurious interactions between the
repeated images.

\begin{figure}[!b]
\begin{center}
\includegraphics[width=0.95\linewidth]{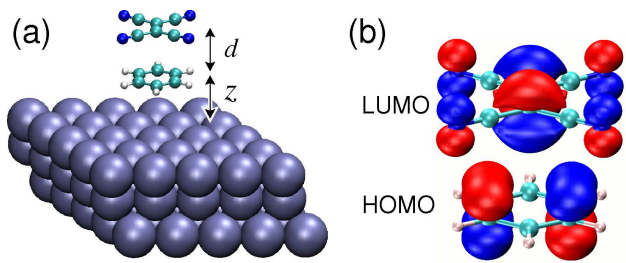}
\caption[system]{\label{fig2} (Color online) (a) Supercell used to describe the
benzene and benzene-TCNE molecules on Al(111). (b) Contour plots of the HOMO and
LUMO orbitals of the benzene-TCNE charge transfer complex.}
\end{center}
\end{figure}

Table I summarizes our results for the four different
donor-TCNE complexes in the gas-phase. The three columns show the
calculated and experimental QP gaps, optical gaps, and electron-hole
interactions. The latter has been defined according to \cite{note2}
\begin{equation}\label{eq.delta_opt}
E_{\text{gap}}^{\text{opt}}=E^{\text{QP}}_{\text{gap}}+E_{\text{e-h}}
\end{equation}
In addition to the many-body results we have performed TDLDA and
$\Delta$SCF(LDA) calculations for comparison. The results obtained
with the two latter methods are in good agreement with another recent
study\cite{ziegler}.

Both BSE and $\Delta$SCF significantly improve the charge-transfer excitations
compared to TDLDA -- in particular the ordering of the gap sizes becomes correct.
The error in the BSE optical gap is due to the
additive effects of an underestimation of the QP gap and
overestimation of the \emph{e-h} interaction with the latter being the largest
effect. In contrast, the $\Delta$SCF method provides an
accurate description of the \emph{e-h} interaction, while the underestimation of the
QP gap is the main source of error. We mention that recent TDDFT calculations
employing a range-separated
functional achieved a mean absolute error (MAE) of only 0.12 eV for the four
molecules by tuning the range parameter separately for each
molecule\cite{stein_kronik}. A recently introduced constrained variational DFT
scheme yields a MAE of around 0.2 eV.\cite{ziegler}

\begin{figure}[!t]
\begin{center}
\includegraphics[width=.9\linewidth]{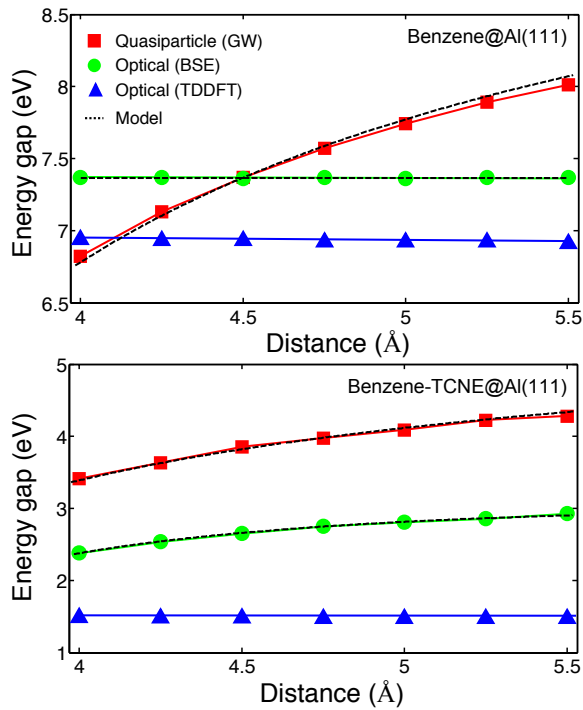}
\caption[system]{\label{fig4} (Color online) Calculated QP and optical gaps of
benzene (upper panel) and benzene-TCNE (lower panel) as a function of distance to
the Al(111) surface.}
\end{center}
\end{figure}

We next consider how the excitation energies are affected when the molecules are
placed next to a metal surface. Here we focus on the regime where hybridization
between the metal and molecule wave functions can be neglected; we have found this
to be the case for $z\geq 4.0$~\AA. In this regime the COHSEX approximation to the
GW self-energy can be used to obtain the change in the QP energy levels\cite{neaton}
\begin{equation}
\Delta E_n^{\text{QP}}=\pm \frac{1}{2} (\psi_n^*\psi_n|\Delta W|\psi_n^*\psi_n)
\end{equation}
where the sign is positive (negative) for empty (occupied) states and we have
introduced the shorthand
\begin{equation}
(f|A|g) = \int \int f(r)A(r,r')g(r')drdr'.
\end{equation}
The change in the QP gap thus becomes
\begin{equation}\label{eq.deltaQP}
\Delta E^{\text{QP}}_{\text{gap}}= \frac{1}{2} [(\psi_H^*\psi_H|\Delta
W|\psi_H^*\psi_H)+(\psi_L^*\psi_L|\Delta W|\psi_L^*\psi_L)]
\end{equation}
The potential $\Delta
W\equiv W_{\text{mol@surf}}-W_{\text{mol}}$, is the change in the (static) screened
Coulomb interaction due to the presence of the surface. Physically it represents the
electrostatic potential at
point $r'$ due to the polarization of the metal induced by a point
charge at point $r$. Clearly this interaction is attractive, i.e. $\Delta W<0$. The
simplest approximation to $\Delta W$ is the classical image potential model
\begin{equation}\label{eq.image}
\Delta W(r,r')=-\frac{1}{2}[(x'-x)^2+(y'-y)^2+(z'+z)^2]^{-1/2}
\end{equation}
which reduces to the well known form $1/4z$ for $r=r'$.

In Fig. \ref{fig4} (red squares) we show the calculated QP gap of
benzene (upper panel) and benzene-TCNE (lower panel) as a function of
the distance to the Al(111) surface. The GW results show the expected
closing of the gap due to the image charge
effect\cite{neaton,thygesen_image,garcia}. The dashed lines show the
result of the classical image charge model Eq. (\ref{eq.image}) where the HOMO and LUMO
charge distributions have been approximated by point charges and the image plane
position has been used as fitting parameter\cite{garcia}.

Within the standard BSE approach, the optical excitations are obtained by
diagonalizing an effective two-particle
Hamiltonian within a space of single-particle
transitions\cite{rubio_RMP}. To simplify the discussion we shall assume that we can
neglect the mixing of single-particle transitions (this in fact turns out to be a
valid approximation). In this case the \emph{e-h} interaction corresponding to the
HOMO-LUMO transition is simply given by the diagonal element of the
exchange-correlation kernel
\begin{equation}\label{eq.e-h}
E_{e-h}=(\psi_H^*\psi_L|v|\psi_L^*\psi_H)-(\psi_H^*\psi_H|W|\psi_L^*\psi_L)
\end{equation}
The first term is a repulsive electron-hole exchange interaction and the second term
is a direct screened Coulomb interaction. The change in the \emph{e-h} interaction
due to the metal surface is then
\begin{equation}\label{eq.deltae-h}
\Delta E_{e-h}=-(\psi_H^*\psi_H|\Delta W|\psi_L^*\psi_L)
\end{equation}
(where we have assumed weak coupling so that the orbitals of the adsorbed molecule
are similar to those in the gas phase.)
Combining Eqs. (\ref{eq.delta_opt}), (\ref{eq.deltaQP}), and (\ref{eq.deltae-h}) we
arrive at the following expression for the change in the optical gap
\begin{equation}\label{eq.delta_opt2}
\Delta E^{\text{opt}}_{\text{gap}}=\frac{1}{2}(\psi_H^*\psi_H-\psi_L^*\psi_L|\Delta
W|\psi_H^*\psi_H-\psi_L^*\psi_L)
\end{equation}
This form is very suggestive showing that the change in optical gap is given by the
transition-generated dipole interacting with its image in the metal surface. Since
$\Delta W<0$ the optical gap is always reduced upon adsorption. We stress that the
static COHSEX and ``diagonal exciton'' approximations
have been used in the discussion above for illustrative purposes only,
and they have not been applied in the \emph{ab-initio} calculations.

In Fig. \ref{fig4} we show the optical gap of the molecules calculated
from the BSE (green circles) and ALDA (blue triangles). The optical gaps correspond
to the $\pi \to \pi^*$ charge-transfer excitation of benzene-TCNE and the
$^1$A$_{1g}$ $\to$ $^1$E$_{1u}$ transition in benzene which is the brightest of the
four HOMO-LUMO transitions. Focusing first
on benzene (upper panel) both calculations yield no change in the
optical gap upon adsorption. This is easily understood from Eq.
(\ref{eq.delta_opt2}) since the dipole moment of the HOMO-LUMO
transition in benzene is negligible. Another way of stating this is
that the change in the QP gap is completely outbalanced by the
weakening of the \emph{e-h} interaction due to screening. This is in good agreement
with experiments on benzene adsorbed on the Ag(111) surface which showed a change in
the optical gap of less than 0.05 eV as compared to benzene in the
gas-phase\cite{avouris}.

Interestingly, for distances $z< 4.5$ \AA, the QP gap is smaller than
the optical gap implying, somewhat counterintuitively, that the
electron-hole interaction is repulsive.  According to Eq.
(\ref{eq.e-h}) this can occur if the screened direct \emph{e-h} interaction
becomes smaller (in absolute value) than the \emph{e-h} exchange energy. From a
gas-phase calculation we have found that the exchange and screened direct \emph{e-h}
interactions are 2.0 eV and -5.1 eV, respectively. While the former is unchanged by
the metal, the latter is reduced due to the image charge effect. Using the fact that
$|\psi_H(r)|$ and $|\psi_L(r)|$ are very similar for benzene, Eqs.
(\ref{eq.deltaQP}) and (\ref{eq.deltae-h}) show that $\Delta E_{e-h}$ and $\Delta
E^{QP}_{gap}$ coincide. Since the QP gap of benzene is 10.5 eV in the gas phase, we
see from Fig. \ref{fig4} that $\Delta E^{QP}_{gap}$= 3.1 eV at $z= 4.5$ \AA. Thus
the exchange and screened direct interactions exactly cancel at this distance. At
even smaller distances, the exchange dominates the screened direct interaction and
the optical gap exceeds the QP gap.

We should stress that the results presented above refer to optical
transitions on a single molecule. Since our calculations
are performed with periodic boundary conditions we are in fact simulating an
infinite array of molecules. Because of the large inter-molecular distance of
$>10$~\AA there is no hybridization between the repeated images.  Formally, however,
the optical gap of the periodic array is always lower than the QP gap since the
electron and hole can be created infinitely far apart in which case the \emph{e-h}
interaction vanishes. Such excitonic states with the electron and hole located on
separate molecules also appear as solutions to the BSE. However, their dipole
strength vanishes and they are thus irrelevant for the low coverage of molecules
considered here.

For the benzene-TCNE complex the BSE calculation predicts a reduction
of the optical gap from its gas-phase value of $3.22$ eV to $2.38$ eV
when the molecule is adsorbed $z=4.0$ \AA~above the metal surface.
The change in the optical gap is very accurately reproduced by an
image dipole model (dashed line) in which $\Delta W$ is again modeled
by a classical image potential and the HOMO and LUMO charge
distributions are modeled as point charges separated by $d=3.6$ \AA.
The classical theory of dipole radiation was used several
decades ago to study the related problem of frequency shifts in
electric dipole emitters by a nearby metal
surface\cite{chance,maniv,holland}. In these studies the role of
the frequency dependence of the metal dielectric function has been
emphasized. On the other hand, recent GW calculations have shown that the
renormalization of the QP levels is well described by the static
response function of the metal\cite{neaton,thygesen_image}. On this
basis it is reasonable to expect that the use of the static part of
$W$ for the \emph{e-h} interaction in the BSE is also a valid approximation.

In summary, we have shown that non-local correlation effects (image charge effects)
reduce the optical excitation energies of a molecule by an amount proportional
to the transition-generated dipole when it is placed near a polarizable medium.
Moreover, electron-hole binding energies can be reduced by several electron volts
and, as a consequence, the optical gap of a molecule can exceed its quasiparticle
gap close to a metal surface.

The authors thank Tom Ziegler and Angel Rubio for inspiring discussions.
We acknowledges support from The Lundbeck Foundation's Center for Atomic-scale
Materials Design and the Danish Center for Scientific Computing.



\begin{thebibliography}{40}
\bibitem{note2}
While this equation can always be taken as the definition of $E_{\text{e-h}}$, its
physical content is most clear in cases where the lowest optical excitation can be
well approximated by the single transition defining the QP gap. In this case all
quantities in Eq. (\ref{eq.delta_opt}) refer only to the HOMO and LUMO orbitals. We
have found that this is indeed the case for all the molecules studied in this work.
\end{thebibliography}
\end{document}